%% file: main.tex
\documentclass[
    aps,pre,superscriptaddress,reprint,
    ]{revtex4-2}

\DeclareUnicodeCharacter{2212}{\textminus}
\usepackage{graphicx} % Required for inserting images

\usepackage[normalem]{ulem}
\usepackage{amsmath}
\usepackage{booktabs}
\usepackage[version=3]{mhchem}
\usepackage{amssymb}
\usepackage{color}

\usepackage[colorinlistoftodos, color=gray!5]{todonotes}
\usepackage{enumitem}
\usepackage{comment}
\usepackage{placeins}
\usepackage[caption=false]{subfig}
\usepackage{hyperref}

\definecolor{green}{RGB}{0, 150, 100}

\begin{document}
\title{Solubility of poorly soluble species in bulk and nanopores from grand-canonical Wang--Landau sampling}
\author{Russell Kajouri}
\affiliation{Institute for Computational Physics, University of Stuttgart, Allmandring 3, 70569 Stuttgart, Germany.}

\author{Adyant Agrawal}\email{aagrawal@icp.uni-stuttgart.de}
\affiliation{Institute for Computational Physics, University of Stuttgart, Allmandring 3, 70569 Stuttgart, Germany.}

\author{Philipp St\"{a}rk}
\affiliation{Institute for Computational Physics, University of Stuttgart, Allmandring 3, 70569 Stuttgart, Germany.}
\affiliation{SC SimTech, Stuttgart Center for Simulation Science, University of Stuttgart, 70569 Stuttgart, Germany.}
\affiliation{Institute for Physics of Functional Materials, Hamburg University of Technology, 21073 Hamburg, Germany.}

\author{Alexander Schlaich}
\affiliation{Institute for Physics of Functional Materials, Hamburg University of Technology, 21073 Hamburg, Germany.}

\author{Christian Holm}\email{holm@icp.uni-stuttgart.de}
\affiliation{Institute for Computational Physics, University of Stuttgart, Allmandring 3, 70569 Stuttgart, Germany.}

\date{\today}

%%%%%%%%%%%%%%%%%%%%%%%%%
%%% Abstract
\begin{abstract}
\input{sections/abstract}
\end{abstract}
\maketitle

%%%% Sections
\section{Introduction}
\input{sections/introduction}

\section{Methods}
\label{sec:methods}

\subsection{Simulation setup}
\input{sections/models}

\subsection{Grand-canonical sampling via Wang--Landau method}
\input{sections/methods}

\section{Results}
\input{sections/results}

\section{Conclusion}
\input{sections/conclusion}

\begin{acknowledgments}
We gratefully acknowledge financial support from the Deutsche Forschungsgemeinschaft (DFG, German Research Foundation)---Project-ID 358283783---SFB 1333/2 2022, as well as funding provided under Germany’s Excellence Strategy—EXC 2075-390740016.
A.S. acknowledges support from the Stuttgart Center for Simulation Science (SimTech).
C.H. further acknowledges computational support from the DFG Compute Cluster funded under Grant No. 261833929 and INST 41/1148-1 Holm 492175459. Additional support was provided by the state of Baden-W\"{u}rttemberg through bwHPC and by DFG under grant INST 35/1597-1 FUGG (bwForCluster Helix). Computational resources and support were also provided by the High Performance and Cloud Computing Group at the Zentrum f\"{u}r Datenverarbeitung, University of T\"{u}bingen, with funding from the state of Baden-W\"{u}rttemberg through bwHPC and from the DFG under Project number 455787709 (bwForCluster BinAC2).
\end{acknowledgments}

\input{sections/appendix}

\bibliography{sections/references}

\end{document}

%% file: sections/abstract.tex
Poorly soluble species can have mean populations below one in small systems such as nanopores.
In this regime, the grand-potential minimum, which identifies the most probable particle number, differs from the ensemble-averaged solute loading.
We use grand-canonical Wang--Landau sampling (GCWLS) to calculate the density of states $\Omega(N_\mathrm{A},N_\mathrm{B})$ for a binary Lennard-Jones mixture of solute B in solvent A, both in bulk and in slit pores.
One set of GCWLS simulations gives ensemble averages over a range of chemical potentials, whereas conventional grand-canonical Monte Carlo (GCMC) simulation requires a separate simulation at each state point.
Independent GCMC simulations confirm the GCWLS ensemble averages.
The solvent number distribution is nearly Gaussian, whereas the dilute solute distribution is skewed and Poisson-like.
The most probable solute number therefore lies below the ensemble-averaged loading $\langle N_\mathrm{B}\rangle$ and is zero whenever $\langle N_\mathrm{B}\rangle<1$.
Reading the solute loading from the grand-potential minimum underestimates it and can falsely indicate that no solute is present.
Defining the oversolubility as $\mathcal{O}_\mathrm{B}=x_\mathrm{B}^{\mathrm{conf}}/x_\mathrm{B}^{\mathrm{bulk}}$, where $x_\mathrm{B}$ is the solute mole fraction, we find $\mathcal{O}_\mathrm{B}\approx12$ in a pore of 3.5 particle diameters width when both wall interactions are repulsive.
Adding wall--solute attraction moderately raises it to $\approx13$, whereas wall--solvent attraction lowers it to $\approx3.4$.
This comparison shows that the wall--solvent interaction controls most of the enhancement.
As the pore widens, $\mathcal{O}_\mathrm{B}$ decreases toward unity.
In bulk and in nanopores alike, the solubility of poorly soluble species should be obtained from ensemble averages rather than from grand-potential minima.

%% file: sections/introduction.tex
The dissolution of low soluble gases, such as \ce{CO2} and \ce{CH4}, in liquids like \ce{H2O} is central to environmental applications, such as gas-hydrate formation\cite{Algaba2023} and \ce{CO2} capture and storage in brines and water-filled porous media\cite{Blazquez2024, Li2020, Du2024, Li2024, Li2025, Niu2026}.
Accurate predictions of gas solubility, phase behaviour, and molecular transport are needed to design carbon-capture\cite{Mathur2026, Micari2025}, catalysis\cite{Zhang2024, Zhang2024_MOF}, and gas-storage\cite{Coasne2019} technologies.
Experiments show that gases can dissolve in liquids confined in mesoporous solids up to about twenty times more than in the bulk liquid\cite{Miachon2008, PeraTitus2009, SoubeyrandLenoir2012, Clauzier2012}.
Molecular simulations attribute such oversolubility to the interactions of both the gas and the solvent with the pore walls\cite{Ho2011, Ho2013, Ho2015, Coasne2019}.
In water-filled pores, the enhancement is generally strongest at hydrophobic walls, an effect attributed to adsorption and clustering of the gas at the surface\cite{Luzar2005, Gadikota2017, Li2020, Xie2026, Choudhary2024}, while hydrophilic walls, covered by strongly bound water layers, enhance the solubility less or even reduce it\cite{Li2020, Xie2026}.
At a hydrophobic wall, however, gas adsorption and weak solvent binding occur together.
More generally, changing the gas or the porous solid alters several of these interactions at once, and reported solubilities for the same gas--liquid pair vary widely between studies\cite{Niu2026}, so the separate contributions of these interactions to the enhancement remain unknown.

Gas solubilities in such systems are usually computed with grand-canonical Monte Carlo (GCMC) simulations\cite{frenkel2023understanding}.
For a poorly soluble gas, however, insertions into the dense solvent are rarely accepted, so the few solute molecules are exchanged too seldom for their average to converge\cite{Mezei1980, Smit1995}.
Cavity-biased\cite{Mezei1980} or configurational-bias\cite{Siepmann1992, Smit1995} insertion, gradual insertion of particles\cite{Ashton2010, Rahbari2020}, and larger cells reduce these effects at higher cost, but the simulation still visits each particle number in proportion to its probability, so rarely visited states remain poorly sampled.
Enhanced-sampling methods have therefore been developed to improve exploration of rarely visited states and transitions between them\cite{Besold1999, Wilding1995}. Transition-Matrix Monte Carlo\cite{ShenErrington2013} (TMMC) and Wang--Landau sampling\cite{Wang2001, Shell2002, desgranges2012evaluation, Stark2026} are two closely related prominent examples: the former reconstructs state probabilities from transition statistics, whereas the latter estimates the density-of-states (DOS) through adaptive biasing. 
In this manuscript, we use grand-canonical Wang--Landau sampling (GCWLS), which yields the two-dimensional DOS, $\Omega(N_\mathrm{A}, N_\mathrm{B})$, and from it the particle-number distribution $\Gamma(N_\mathrm{A}, N_\mathrm{B})$ over a range of chemical potentials, from a single set of simulations.
As our analysis relies only on $\Gamma$, it is independent of the sampling algorithm.

Grand-potential profiles in the particle number, $\Phi = -k_\mathrm{B}T\ln\Gamma$, are commonly interpreted through their minima, which identify the stable states, for instance the filled and empty states of water in nanopores\cite{Hummer2001, Rasaiah2008}.
For a solute present at only a few molecules, however, the statistics of small open systems depart from macroscopic expectations\cite{Neimark2005, Dixit2017, Wilding2026}, and the minimum of $\Phi$ need not coincide with the equilibrium uptake.
Such low occupancies are common, since they arise from poor solubility, from low gas pressure, and from the small volume of a nanopore.

Here, we determine how the minimum of $\Phi$ relates to the equilibrium uptake of a solute present at only a few molecules and how the wall--solvent and wall--solute interactions separately contribute to the oversolubility.
We use GCWLS for a binary Lennard-Jones mixture composed of a low-solubility solute B and a solvent A, both in bulk and under slit-pore confinement.
For the bulk system, we compare the minimum of $\Phi$ with the equilibrium ensemble-average solute loading, $\langle N_\mathrm{B}\rangle$.
For the confined system, we isolate the role of wall--fluid interactions by switching off the wall attraction for one component at a time: (I) the walls repel both components, (II) the walls repel only the solvent, and (III) the walls repel only the solute.
We examine each case at three pore widths that span the range from strong to weak confinement.

%% file: sections/models.tex
\begin{figure*}[ht]
    \centering
    \includegraphics[width=\linewidth]{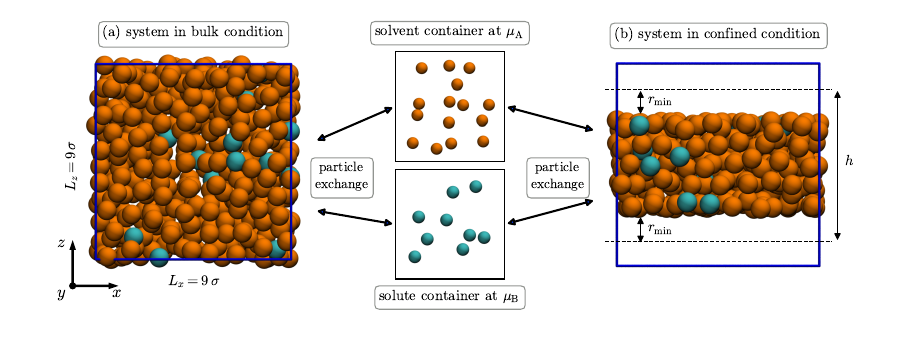}
    \caption{The schematic picture of the system is illustrated in bulk (a) and confined (b) conditions, where the both systems are in touch with distinct ideal solvent and solute containers at chemical potentials $\mu_{\mathrm{A}}$ and $\mu_{\mathrm{B}}$, respectively. In panel (b), the dashed black lines denote the wall centre-of-mass.}
    \label{fig:system}
\end{figure*}

We perform simulations of a binary mixture of poorly soluble species B (solute) in abundant species A (solvent) in both bulk and confinement, as illustrated in Fig.~\ref{fig:system}(a) and (b), respectively. 
All particles have the same mass $m$ and diameter $\sigma$, which, together with the solvent--solvent well depth $\epsilon \equiv \epsilon_{\mathrm{AA}}$, constitute the units of mass, length, and energy. 
All quantities hereafter are reported in these Lennard-Jones (LJ) reduced units~\cite{frenkel2023understanding}. 
Particles interact through a truncated and shifted Lennard-Jones (12-6) potential,
\begin{equation}
    U_{\mathrm{LJ}}(r) = 4\epsilon_{ij}\left[(\sigma_{ij}/r)^{12} -(\sigma_{ij}/r)^{6}\right]
\end{equation}
for $r < r_c = 2.5\sigma$ and shifted-to-zero beyond, where $r = \lVert \mathbf{r}_i - \mathbf{r}_j \rVert$ denotes the distance between particles $i,j \in \{\mathrm{A,B}\}$, and $\epsilon_{ij}$ and $\sigma_{ij}$ represent the potential well depth and effective diameter for the $i$–$j$ pair of components in the solution.
The solute--solute interaction is weaker than that of the solvent--solvent, $\epsilon_{\mathrm{AA}} = 1\epsilon$ and $\epsilon_{\mathrm{BB}} = 0.1\epsilon$. This choice places B in the low-solubility regime. 
Cross interactions follow the Lorentz-Berthelot rules\cite{allen2017chapter1}, giving the ordering $\epsilon_{\mathrm{AA}} > \epsilon_{\mathrm{AB}} > \epsilon_{\mathrm{BB}}$.

In the bulk simulations, we employ a cubic simulation box with 
an edge length of $L = 9\,\sigma$ and 
impose periodic boundary conditions in all three spatial directions. 
In contrast, in the confined geometry, the slit pore is represented by two infinitely extended, parallel walls lying in the $x$--$y$ plane and separated by a distance $h$, along the $z$ axis (Fig.~\ref{fig:system}(b)).
The simulation cell has lateral dimensions identical to the bulk box, $L_x = L_y = L$, with periodic boundary conditions in the $x$ and $y$ directions, whereas along $z$ the system is confined and non-periodic.
The interactions between the confining walls and the fluid particles are described by a Lennard-Jones (9-3) potential, expressed as
\begin{equation}
    \label{eq:LJ93}
    U(r) = \epsilon_{\mathrm{W}i}
    \Bigg[ \frac{2}{15}
        \Big( \frac{\sigma_{\mathrm{W}i}}{r}\Big)^9 - 
        \Big( \frac{\sigma_{\mathrm{W}i}}{r}\Big)^3
    \Bigg],\quad r<r_{c,\mathrm{W}},
\end{equation}
Here, $r$ denotes the separation between solution particles and the confining walls measured along the surface-normal direction.
The two wall types share the same repulsive core and differ only in whether the attractive tail is retained. 
In the purely repulsive form, the LJ(9-3) potential is truncated at its minimum, $r_{c,\mathrm{W}} = r_{\mathrm{min}} = (2/5)^{1/6}\,\sigma$, and shifted-to-zero, so that only repulsion remains (the Weeks--Chandler--Andersen construction~\cite{WCA1971role}).
In the full Lennard-Jones form, the attractive tail is retained and the potential is truncated and shifted-to-zero at $r_{c,\mathrm{W}} = 5\sigma$. 
The energy interaction parameters between the confining walls and the components are taken to be equal to the corresponding self-interaction values, that is, $\epsilon_{\mathrm{WA}}=\epsilon_{\mathrm{AA}}$ and $\epsilon_{\mathrm{WB}}=\epsilon_{\mathrm{BB}}$. The effective pore width is defined for both types of wall--solution interactions as \( h_{\mathrm{eff}} := h - 2r_{\mathrm{min}} \).

%% file: sections/methods.tex
We study the system primarily with grand-canonical Wang--Landau sampling (GCWLS) and employ grand-canonical Monte-Carlo (GCMC) as an independent cross-check. Both operate in the grand canonical ensemble $\left(\mu_A,\mu_B,V,T\right)$ at fixed temperature $T = 1$ (in units of $\epsilon/k_B$, with $k_B$ the Boltzmann constant) and fixed volume, $L^3$ for the bulk box and $L^2\,h$ for the slit pore with $L = 9\,\sigma$ (see Fig.~\ref{fig:system}).
We keep the solvent chemical potential at $\mu_A = -1.1\epsilon$, which places the solvent in the liquid phase away from vapour--liquid coexistence, and vary the solute chemical potential $\mu_B$ to trace the solubility.

The two methods differ in their treatment of the chemical potential $\mu$. In GCMC, $\mu$ is prescribed during sampling, requiring a separate simulation for each value. In GCWLS, the density-of-states $\Omega$ is sampled independently of $\mu$, enabling the equilibrium behaviour over the entire $\mu_B$ range to be recovered through post-processing.
Detailed balance fixes the ratio of the exchange (insertion/deletion) acceptance probabilities, common to both methods and differing only in the weight $w$,
\begin{widetext}
\begin{equation}
    \label{eq:acceptance}
    \frac{\mathrm{acc}(N \rightarrow N{+}1)}{\mathrm{acc}(N{+}1 \rightarrow N)}
    = \frac{V}{\Lambda^3 (N{+}1)}\, w \,
      \exp\!\left[-\beta\left(\mathcal{H}(\mathbf{r}^{N+1})
      - \mathcal{H}(\mathbf{r}^{N})\right)\right],
    \qquad
    w =
    \begin{cases}
        \exp(\beta\mu) & \text{(GCMC)},\\[3pt]
        \dfrac{\Omega(N)}{\Omega(N{+}1)} & \text{(GCWLS)},
    \end{cases}
\end{equation}
\end{widetext}
where $N$ is the number of particles of the exchanged species with the other held fixed, $\Lambda $ is the de-Broglie thermal wavelength (set to unity), $\beta = 1/(k_B T)$, and $\mathcal{H}(\mathbf{r}^{N})$ is the configurational (potential) energy of the $N$-particle state.
GCWLS estimates the configurational density-of-states, $\Omega(N_A,N_B)$, associated with each particle-number macrostate. The grand partition function is then obtained by weighting these macrostates according to the imposed chemical potentials and summing over all values of $N_A$ and $N_B$, as described mathematically in Eq.~\eqref{eq:grand_partition_function}, 

\begin{equation}
    \label{eq:grand_partition_function}
    \Xi = \sum_{N_{A}, N_{B}}
    \Omega(N_A,N_B)\, \exp\!\left[\beta\left(N_{A}\mu_{A} + N_{B}\mu_{B}\right)\right],
\end{equation}
and the joint particle-number distribution, i.e.\ the grand-canonical probability of finding $N_A$ solvent and $N_B$ solute particles,
\begin{equation}
    \label{eq:joint_PND}
    \Gamma(N_A,N_B) = \frac{\Omega(N_A,N_B)}{\Xi}\,
    \exp\!\left[\beta\left(N_{A}\mu_{A} + N_{B}\mu_{B}\right)\right],
\end{equation}
for any set of chemical potentials~\cite{desgranges2012evaluation,ganzenmuller2007applications,Shell2002}.
Since GCMC spends most of its time in the high-probability states near the grand-potential minimum, the low- and high-$N$ tails of the distribution $\Gamma$ (Eq.~\eqref{eq:joint_PND}) are sampled sparsely.
Nevertheless, a sufficiently long Monte Carlo simulation can construct a normalised histogram of states near equilibrium, providing an empirical estimate of $\Gamma$ in that region.
However, GCWLS removes this limitation as, under its flat-histogram bias, it visits every macrostate $(N_{\mathrm{A}}, N_{\mathrm{B}})$ equally. To build this bias, $\Omega$ is refined iteratively: after each attempted exchange, accepted or rejected (Eq.~\eqref{eq:acceptance}), the density-of-states of the macrostate $(N_A, N_B)$ is updated according to
\begin{equation}
    \label{eq:Q_refining_rules}
    \Omega_{\mathrm{new}}(N_A, N_B) = f(k)\,\Omega_{\mathrm{old}}(N_A, N_B),
\end{equation}
with the modification factor $f(k) = \exp(\alpha/2^{k}) > 1$ decreasing toward unity over the simulation stages $k \in \{0,1,2,\dots\}$.
A histogram records how often each macrostate is visited. Once every macrostate has been visited at least $N_v = \varepsilon/\sqrt{\ln f}$ times, it is reset and the stage $k$ advanced, following Zhou and Bhatt~\cite{Zhou2005} (here $\alpha = 1$, $\varepsilon = 1000$).
Once $\Omega$ has converged, $\Xi$ and $\Gamma$ (Eqs~\eqref{eq:grand_partition_function} and~\eqref{eq:joint_PND}) are evaluated over the full range of chemical potentials.

As in Ref.~\cite{Stark2026}, separate simulations vary either $N_{\mathrm{A}}$ or $N_{\mathrm{B}}$ within a sampling window while the other is held fixed, and the resulting data are combined in post-processing to reconstruct the two-dimensional density of states $\Omega(N_{\mathrm{A}}, N_{\mathrm{B}})$.
The sampled range of macrostates is partitioned into fixed-size windows, each containing five distinct macrostates, with three macrostates shared between adjacent windows. The resulting $\Omega$ profiles are subsequently stitched together by minimizing the root-mean-square deviation (RMSD) within the overlapping regions; the corresponding error $\sigma_{\mathrm{stitching}}$ is defined in Appendix~\ref{app:errors}.
An additional uncertainty stems from the modification factor ($f$) employed in the Wang--Landau iteration, which induces a statistical error in the estimate of $\Omega$ that scales as $\sqrt{\ln f(k,\alpha)}$~\cite{Zhou2005}.
To compare this scaling factor with the density of states of our system, after removal of the arbitrary additive offset, the root-mean-square change $D_k$ in $\Omega$ over each two successive iterations is calculated, resulting in $D_{19}=O(10^{-1})$ for the last two iterations (Fig.~\ref{fig:RMS_variation_lnOmega}). Its definition is given in Appendix~\ref{app:errors}.
The error $\sigma_{\ln\Omega}$ of the density of states associated with each macrostate $\mathbf{N}=(N_\mathrm{A},N_\mathrm{B})$ is then defined as
\begin{equation}
    \sigma_{\ln\Omega}(\mathbf{N}) = \sqrt{\sigma^2_{\mathrm{stitching}}(\mathbf{N}) + D^2_{k_{\max}}},
\end{equation}
where $D_{k_{\max}}$ is the value of $D_k$ at the final iteration and the stitching error $\sigma_{\mathrm{stitching}}$ measures the residual disagreement among overlapping DOS windows after the global
least-squares alignment (Appendix~\ref{app:errors}).

The marginal distribution of species $i$ is obtained by summing the joint distribution over the particle number of the other species, $\Gamma_i(N_i) = \sum_{N_j}\Gamma(N_i,N_j)$, with $i,j\in\{A,B\}$ and $i\neq j$.
The corresponding marginal grand-potential profile is
\begin{equation}
    \Phi_i(N_i) = -k_{\mathrm{B}}T \ln \Gamma_i(N_i).
    \label{eq:Phi_from_joint_PND}
\end{equation}
Although the grand-potential profile minimum identifies the most probable value, equilibrium composition and particle-number fluctuations are determined by the ensemble average (mean) and variance of $\Gamma$, respectively~\cite{desgranges2012evaluationI}:
\begin{align}
    \langle N_i\rangle &= \sum_{N_A}\sum_{N_B} N_i\,\Gamma(N_A,N_B),
    \label{eq:ensemble_average}\\
    \mathrm{Var}(N_i) &= \sum_{N_A}\sum_{N_B}
    \left(N_i - \langle N_i\rangle\right)^2 \Gamma(N_A,N_B),
    \label{eq:variance}
\end{align}
with $i\in\{\mathrm{A},\mathrm{B}\}$.
The solute mole fraction is evaluated for both the bulk and confined systems as
\begin{equation}
    x_{\mathrm{B}} = \frac{\langle N_\mathrm{B}\rangle}{\langle N_\mathrm{A}\rangle + \langle N_\mathrm{B}\rangle}.
\end{equation}

The solute oversolubility is then quantified by the ratio of the confined solute mole fraction to its bulk counterpart:
\begin{equation}
    \mathcal{O}_{\mathrm{B}} = {x^{\mathrm{conf}}_\mathrm{B} } / {x^{\mathrm{bulk}}_\mathrm{B}}.
    \label{eq:solute_over_solubility}
\end{equation}
The errors of $x_\mathrm{B}$ and $\mathcal{O}_\mathrm{B}$ are obtained by first-order propagation of the density-of-states error $\sigma_{\ln\Omega}$ (Appendix~\ref{app:errors}).

To sample the configurational phase space, both Monte Carlo schemes are interleaved with molecular-dynamics blocks, each comprising $5\times10^3$ integration steps with a time step of $\delta t=0.005,\tau$, where $\tau=\sqrt{m\sigma^2/\epsilon}$ denotes the characteristic MD time unit. For both GCMC and GCWLS, each Monte Carlo block consists of $5\times10^2$ steps. GCMC simulations are sampled over a total of $5\times10^7$ MC steps, including insertion/deletion. For convergence  of the density-of-states in GCWLS $k_{\max}=19$ resets (iterations) is sufficient.
The system temperature is controlled via a Nos\'{e}--Hoover thermostat with a damping timescale of $100\,\delta t$. The equations of motion are integrated via the velocity-Verlet scheme. The simulations are run using LAMMPS(29-Aug-2024)~\cite{LAMMPS,Plimpton1995},
modified to implement grand-canonical Wang--Landau sampling \cite{Philipp2025,Stark2026}. All LAMMPS input files, analysis files are openly available on DaRUS\cite{DaRUS}.

%% file: sections/results.tex
\begin{figure*}
    \centering
    \includegraphics[scale=1.0]{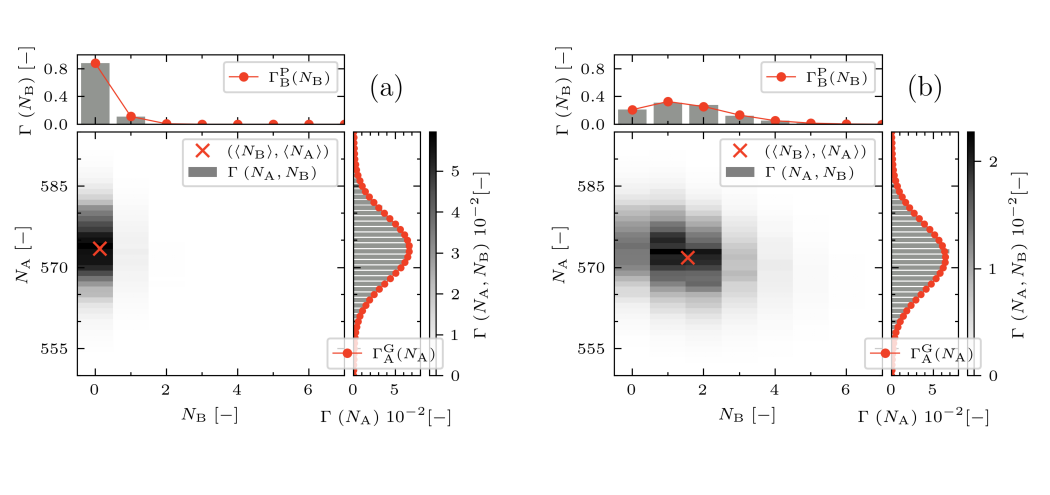}
    \caption{Bulk joint particle--number distribution $\Gamma(N_\mathrm{A}, N_\mathrm{B})$, shown as a heat map with the solute on the horizontal axis and the solvent on the vertical axis. The solute chemical potential increases from (a) $\mu_\mathrm{B} = -3.8\,\epsilon$ to (b) $\mu_\mathrm{B} = -1.5\,\epsilon$, at fixed solvent chemical potential $\mu_\mathrm{A} = -1.1\,\epsilon$. The top and right panels give the solute and solvent marginals, $\Gamma_\mathrm{B}$ and $\Gamma_\mathrm{A}$. The red curves are the reference distributions of Eqs.~\eqref{eq:poisson_reference} and~\eqref{eq:gaussian_reference}, Poisson for the solute and Gaussian for the solvent, each constructed from the measured ensemble mean and variance.}
   \label{fig:bulk_system}
\end{figure*}

In this section, we first examine the bulk mixture to quantify how far the most probable particle number departs from the equilibrium one when only a few poorly soluble solute particles are present in the solution.
In that case, the solute particle-number distribution is asymmetric and well described by a Poisson distribution, whereas the abundant solvent distribution is effectively continuous, symmetric, and well approximated by a Gaussian.
Owing to this asymmetry, the most probable solute number lies below the ensemble average, which gives the equilibrium particle number in the grand-canonical ensemble; independent GCMC simulations confirm the GCWLS averages.
With this established, we use GCWLS ensemble averages for the confined mixture to determine the solute mole fraction and oversolubility, and thus the wall–fluid interaction effect.

\subsection{Bulk solution}
Using GCWLS, the density-of-states $\Omega(N_\mathrm{A}, N_\mathrm{B})$ for the bulk solution is sampled over $N_\mathrm{A} \in [550, 600]$ and $N_\mathrm{B} \in [0, 7]$.
These ranges are chosen to be wide enough that, for the chemical potentials considered here, each particle-number distribution vanishes at its edges (except at the solute's physical bound $N_\mathrm{B} = 0$).
Figure~\ref{fig:bulk_system} shows the joint distribution $\Gamma(N_\mathrm{A}, N_\mathrm{B})$ at two solute chemical potentials, $\mu_\mathrm{B} = -3.8\epsilon$ (a) and $-1.5\epsilon$ (b), both at the same solvent chemical potential $\mu_\mathrm{A} = -1.1\epsilon$.
In both cases, the distribution occupies a single compact region that is broad along $N_\mathrm{A}$ but narrow in $N_\mathrm{B}$, spanning tens of solvent particles but only a few solute particles.
The two marginal distributions (side panels in Fig.~\ref{fig:bulk_system}a,b) have distinct forms, and in both cases their expected shapes are governed by the statistics of the ensemble rather than by fitting. A species dilute enough to be effectively ideal has statistically independent insertion events, so its particle number should follow a Poisson distribution\cite[Chap~12, Sec.~114]{Landau1980StatisticalPhysics}.
\begin{equation}
    \label{eq:poisson_reference}
    \Gamma^{\mathrm{P}}_\mathrm{B}(N_\mathrm{B}) =
    \frac{\langle N_\mathrm{B}\rangle^{N_\mathrm{B}}}{N_\mathrm{B}!}\,
    \exp\!\big(-\langle N_\mathrm{B}\rangle\big),
\end{equation}
whose only parameter is the measured mean $\langle N_\mathrm{B}\rangle$.
In contrast, fluctuations in the abundant solvent are concentrated around its equilibrium macrostate.
In this regime, the distribution is sharply peaked and approaches a Gaussian fluctuation approximation,~\cite[Chap.~12, Sec.~110]{Landau1980StatisticalPhysics}.
\begin{equation}
\label{eq:gaussian_reference}
\Gamma^{\rm G}_{\rm A}(N_\mathrm{A})
\simeq
\frac{1}{\sqrt{2\pi\mathrm{Var}(N_\mathrm{A})}}
\exp\left(
-\frac{(N_\mathrm{A}-\langle N_\mathrm{A}\rangle)^{2}}
{2\mathrm{Var}(N_\mathrm{A})}
\right).
\end{equation}
We find that both Eqs.~\ref{eq:poisson_reference} and \eqref{eq:gaussian_reference} agree well with the measured marginals, $\Gamma_\mathrm{B} \simeq \Gamma^{\mathrm{P}}_\mathrm{B}$ and $\Gamma_\mathrm{A} \simeq \Gamma^{\mathrm{G}}_\mathrm{A}$, over the whole sampled range (red curves and side panels in Fig.~\ref{fig:bulk_system}~a,b).
Increasing $\mu_\mathrm{B}$ from $-3.8\epsilon$ to $-1.5\epsilon$ raises 
$\langle N_\mathrm{B}\rangle$ from $\approx 0.2$ to $\approx 1.6$ and lowers the Poisson skewness $\langle N_\mathrm{B}\rangle^{-1/2}$ accordingly, so the distribution becomes visibly less asymmetric.
At these occupancies it nonetheless remains far from the symmetric limit, which is reached only for $\langle N_\mathrm{B}\rangle \gg 1$.
In the pore, the same increase in $\mu_\mathrm{B}$ raises $\langle N_\mathrm{B}\rangle$ further, and the solute distribution approaches a Gaussian (Fig.~\ref{fig:confined_system} in Appendix~\ref{app:wider}).

\begin{figure}
    \centering
    \includegraphics[width=\linewidth]{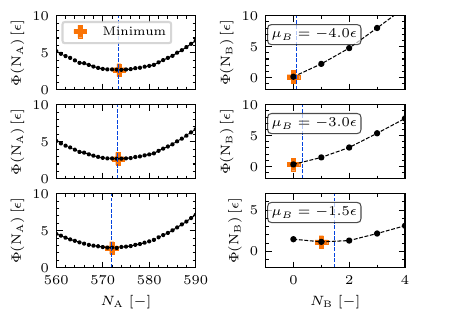}
    \caption{Marginal grand-potential profiles $\Phi$ of the solute (right panels) and solvent (left panels) in the bulk box ($L = 9\,\sigma$), at solvent chemical potential $\mu_\mathrm{A} = -1.1\,\epsilon$ and three solute chemical potentials $\mu_\mathrm{B}$. Orange plus signs mark the minimum of each profile, and blue dashed vertical lines mark the ensemble-averaged particle number $\langle N\rangle$.}
    \label{fig:bulk_phi_L9_GCWLS}
\end{figure}

\begin{table}[!htb]
\centering
\setlength{\tabcolsep}{6pt}
\begin{tabular}{ccccc}
\toprule
& \multicolumn{2}{c}{GCMC}
& \multicolumn{2}{c}{GCWLS} \\
\cmidrule(lr){2-3}\cmidrule(lr){4-5}
$\mu_\mathrm{B}/\epsilon$
& $\langle N_\mathrm{B} \rangle$
& $\mathrm{Var}(N_\mathrm{B})$
& $\langle N_\mathrm{B} \rangle$
& $\mathrm{Var}(N_\mathrm{B})$ \\
\midrule
-3.8 & 0.15 & 0.151 & 0.15 & 0.153 \\
-3.6 & 0.19 & 0.194 & 0.18 & 0.188 \\
-3.4 & 0.22 & 0.224 & 0.22 & 0.232 \\
-3.2 & 0.28 & 0.280 & 0.28 & 0.284 \\
-3.0 & 0.33 & 0.337 & 0.34 & 0.351 \\
-2.8 & 0.42 & 0.419 & 0.42 & 0.431 \\
-2.6 & 0.48 & 0.472 & 0.51 & 0.528 \\
\bottomrule
\end{tabular}
\caption{Mean and variance of the solute particle number in the bulk ($L = 9\,\sigma$, $\mu_\mathrm{A} = -1.1\,\epsilon$) from GCMC and GCWLS. For a Poisson distribution, $\mathrm{Var}(N_\mathrm{B})=\langle N_\mathrm{B} \rangle$.}
\label{tab:solute_mean_variance}
\end{table}

We validate the GCWLS results against independent GCMC simulations performed at fixed $(\mu_\mathrm{A},\mu_\mathrm{B},V,T)$. For the solute, the mean and variance obtained from the two methods agree closely (Table~\ref{tab:solute_mean_variance}), providing an independent validation of the GCWLS implementation. An additional consistency check is provided by the particle-number statistics of the dilute solute. For a Poisson distribution, Eq.~\eqref{eq:poisson_reference}, 
the variance is equal to the mean,
\begin{equation}
    \operatorname{Var}\!\left(N_{\mathrm{B}}\right)=\left\langle N_{\mathrm{B}}\right\rangle .
    \label{eq:poisson_var_equals_mean}
\end{equation}
We indeed find $\mathrm{Var}(N_\mathrm{B})\simeq\langle N_\mathrm{B}\rangle$,\ Table~\ref{tab:solute_mean_variance}.
This behaviour is expected in the dilute limit, where correlations between solute particles are weak and particle-number fluctuations approach those of statistically independent particles in the grand-canonical ensemble. The simultaneous reproduction of this Poisson-like fluctuation behaviour by both GCWLS and conventional GCMC therefore provides a more stringent validation than agreement of the mean population alone.

The distinction between the grand-potential minimum and the ensemble average of $\Gamma$ is most apparent in the marginal grand potential, $\Phi_i=-k_\mathrm{B}T\ln\Gamma_i$ (Fig.~\ref{fig:bulk_phi_L9_GCWLS}). In the grand-canonical ensemble, the equilibrium particle number is represented by the ensemble average $\langle N_i\rangle$, whereas the minimum of $\Phi_i$ identifies the most probable particle number.

For the solvent (left panels), $N_\mathrm{A}$ spans a broad range and $\Phi_\mathrm{A}$ forms a smooth, nearly parabolic well. Consequently, the particle number associated with the grand-potential minimum closely coincides with the ensemble average $\langle N_\mathrm{A}\rangle$. For the dilute solute (right panels), however, only a few discrete values of $N_\mathrm{B}$ are appreciably populated and $\Phi_\mathrm{B}$ is strongly asymmetric. As a result, the particle number at the grand-potential minimum lies below the equilibrium expectation value $\langle N_\mathrm{B}\rangle$ for all $\mu_\mathrm{B}$ explored here. For example, at $\mu_\mathrm{B}=-4\epsilon$, the minimum occurs at $N_\mathrm{B}=0$, while $\langle N_\mathrm{B}\rangle=0.15$; at $\mu_\mathrm{B}=-1.5\epsilon$, it occurs at $N_\mathrm{B}=1$, while $\langle N_\mathrm{B}\rangle\approx1.6$ (see also Fig.~\ref{fig:confined_system} for the confined case).

The consequence is particularly important in the very dilute regime. When $\langle N_\mathrm{B}\rangle<1$, the grand-potential minimum can remain at $N_\mathrm{B}=0$ even though the equilibrium expectation value is finite. Interpreting the location of this minimum as the equilibrium particle number would therefore incorrectly suggest the absence of dissolved solute. Thus, for low-solubility species, the grand-potential minimum should not in general be identified with the equilibrium particle number; the appropriate quantity is the ensemble average value obtained from the particle-number distribution.

As the solute population increases, the distribution becomes less skewed and the distinction between the grand-potential minimum and $\langle N_\mathrm{B}\rangle$ progressively diminishes. Consistent with the approximately Poisson solute statistics, whose skewness scales as $\langle N_\mathrm{B}\rangle^{-1/2}$, the two quantities approach one another as $\langle N_\mathrm{B}\rangle$ increases. For $h=8\sigma$ in case II, for example, the particle number associated with the grand-potential minimum provides a reliable approximation to the equilibrium ensemble average value once $\langle N_\mathrm{B}\rangle\gtrsim4.5$ (Fig.~\ref{fig:confined_system}b).

\subsection{Confined solution}

\begin{table}
    \centering
    \caption{The three confinement cases. Each wall acts on the solvent (A) and the solute (B) through one of the two forms of Eq.~\eqref{eq:LJ93}, either repulsive (WCA) or attractive (full Lennard-Jones). Cases I and II, whose walls repel the solvent, are termed solvophobic, and case III, whose wall attracts it, solvophilic.}
    \label{tab:cases}
    \begin{tabular}{@{}c c c@{}}
        \toprule
        Case & Wall--solvent (A) & Wall--solute (B) \\
        \midrule
        I   & repulsive  & repulsive  \\
        II  & repulsive  & attractive \\
        III & attractive & repulsive  \\
        \bottomrule
    \end{tabular}
\end{table}
\begin{figure}
    \centering
    % subfigure 1
    \subfloat[A slit pore of width $h = 3.5\,\sigma$.\label{fig:mf_3.5}]{
        \includegraphics[width=0.9\linewidth]{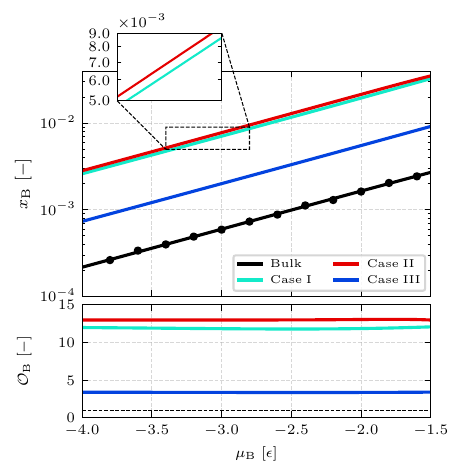}
    }\\ % <--- 
    % subfigure 2
    \subfloat[Mean solute oversolubility\label{fig:average_solute_solubilities}]{
        \includegraphics[width=0.9\linewidth]{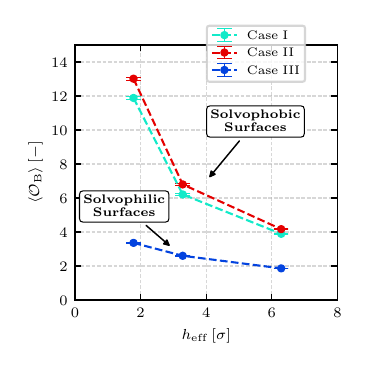}
    }\\
    \caption{(a) Dependence of the solute chemical potential, $\mu_\mathrm{B}$, on the solute mole fraction (upper sub-panel) and on the solute oversolubility under bulk and confined conditions (Cases I–III) for a pore width of $h=3.5\sigma$. Bulk results obtained using the grand-canonical Monte-Carlo (GCMC) method are shown by black symbols, whereas the coloured solid curves correspond to grand-canonical Wang–Landau sampling (GCWLS) calculations. (b) Mean solute oversolubility, $\langle \mathcal{O}_\mathrm{B} \rangle$, as a function of pore width $h_{\mathrm{eff}}$ for the three cases listed in Table~\ref{tab:cases}. The solvophobic wall cases (I and II) yield substantially higher $\langle \mathcal{O}_\mathrm{B} \rangle$ than the solvophilic wall case (III) across the range of pore widths considered.}
\end{figure}

Subsequently, we examine the effects of confinement. For confined geometries, GCWLS is used to compute the density of states in slit pores of widths $h = 3.5$, $5$, and $8\,\sigma$ for each of the three interaction scenarios summarized in Table~\ref{tab:cases}.
An example of the corresponding confined joint distribution is shown in Fig.~\ref{fig:confined_system}, representing case II at a pore of width $h = 8\,\sigma$.
The solute particle number in the pore is small, so it follows a skewed Poisson distribution, and the confined mole fractions $x_{\rm B}$ are obtained from the ensemble averages rather than from the grand-potential minimum.
Confinement-enhanced solubility of poorly soluble species has been reported for many nanoporous materials~\cite{Xie2026, Coasne2019, Li2025, Li2024}, and the three wall cases of Table~\ref{tab:cases} isolate the generic wall-affinity mechanisms responsible rather than modelling a specific material.

Figure~\ref{fig:mf_3.5} (top panel) shows the resulting solute mole fraction, $x_\mathrm{B}$, in the pore. Since one GCWLS run yields $\Omega(N_\mathrm{A}, N_\mathrm{B})$, its full $\mu_\mathrm{B}$ dependence follows from a single sampling run per pore, whereas GCMC would require a separate run at each $\mu_\mathrm{B}$. 
The two solvophobic walls (cases I and II) yield nearly identical (c.f.\ inset in Fig.~\ref{fig:mf_3.5}), strongly enhanced oversolubility with $x_\mathrm{B}$ being roughly an order of magnitude above the bulk value at the same $\mu_\mathrm{B}$, despite their differing wall-solute interactions. 
In contrast, the solvophilic wall (case III) gives a substantially lower, yet about $3.5$ times increased $x_\mathrm{B}$ compared to bulk.

The solute oversolubility $\mathcal{O}_\mathrm{B}$, Eq.~\eqref{eq:solute_over_solubility}, measures the enhancement relative to bulk (Fig.~\ref{fig:mf_3.5}, bottom panel). Across the sampled $\mu_\mathrm{B}$ range, $\mathcal{O}_\mathrm{B}$ is nearly constant, here and in the wider pores (Fig.~\ref{fig:solute_mole_fraction}), because the near-ideal solute has $x_\mathrm{B} \propto e^{\beta\mu_\mathrm{B}}$ in both the pore and the bulk (the log-linear curves in Fig.~\ref{fig:mf_3.5}, top panels), so the $\mu_\mathrm{B}$ dependence cancels in their ratio.
A single $\mu_\mathrm{B}$-averaged value $\langle \mathcal{O}_\mathrm{B} \rangle$ therefore characterises each wall type and pore width (Fig.~\ref{fig:average_solute_solubilities}), reaching $\langle \mathcal{O}_\mathrm{B} \rangle \approx 12$ and $\approx 13$ for the solvophobic walls (cases I and II, respectively) and $\approx 3.4$ for the solvophilic wall (case III) at $h = 3.5\,\sigma$.
Importantly, the small difference between cases I and II shows that the solute excess in the pore is governed by exclusion of the solvent rather than by solute--wall attraction, so the wall--solvent interaction is the dominant factor controlling confinement-enhanced solubility. A wider pore adds an increasingly bulk-like interior that dilutes this excess, so $\langle \mathcal{O}_\mathrm{B} \rangle$ decreases with $h_{\rm eff}$ and tends to unity as the pore widens (Fig.~\ref{fig:average_solute_solubilities}).

\begin{figure}
    \centering
    % subfigure 1
    \subfloat[A slit pore of width $h = 3.5\,\sigma$.\label{fig:subim1}]{%
        \includegraphics[width=0.97\linewidth]{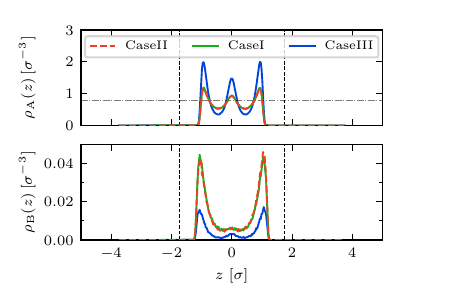}%
    }

    \medskip

    % subfigure 2
    \subfloat[A slit pore of width $h = 5\,\sigma$.\label{fig:subim2}]{%
        \includegraphics[width=0.97\linewidth]{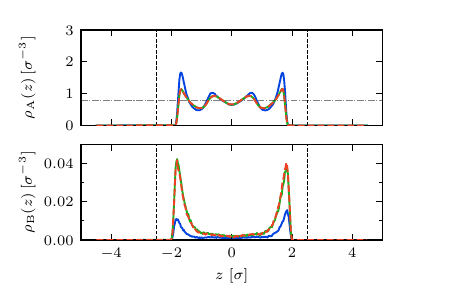}%
    }

    \medskip

    % subfigure 3
    \subfloat[A slit pore of width $h = 8\,\sigma$.\label{fig:subim3}]{%
        \includegraphics[width=0.97\linewidth]{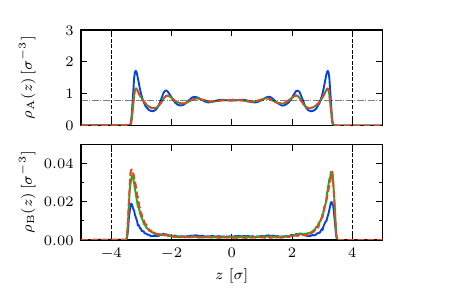}%
    }

    \caption{Solvent density profile $\rho_{\mathrm{A}}$ (top panels) and the solute density profile $\rho_{\mathrm{B}}$ (bottom panels) along the pore coordinate $z$ for three different pore widths. In each panel, the vertical dashed lines represent the centre of mass of the confining walls, and the horizontal dash-dot line represents the bulk solvent density.}
    \label{fig:solvent_density_profile}
\end{figure}

At the equilibrium condition, the density profiles, shown in Fig.~\ref{fig:solvent_density_profile}, identify the spatial origin of the oversolubility, where the density profiles exhibit pronounced oscillations, with the largest amplitude adjacent to each wall. 
In the widest pore $h=8\sigma$, the solvent density returns to its bulk value at the centre (indicated by a horizontal black dash-dotted line), confirming that the pore interior becomes increasingly bulk-like as $h$ increases. In all three cases, the solute accumulates near the wall and is depleted from the interior, so the excess is confined to the near-wall layers.
The solvophobic walls (cases I and II) have nearly coincident profiles and the largest solute excess, whereas the solvophilic wall (case III) shows the strongest solvent layering but a lower solute excess, consistent with the behaviour in $\langle \mathcal{O}_\mathrm{B} \rangle$.

%% file: sections/conclusion.tex
The Grand-canonical Wang--Landau sampling method reconstructs the density-of-states $\Omega(N_\mathrm{A}, N_\mathrm{B})$ of a poorly soluble solute and its solvent from a single set of simulations, so that the full particle-number distribution $\Gamma$ and the ensemble-averaged composition can be computed for the whole sampled range of chemical potentials, both in bulk and under confinement.
Conventional GCMC requires a separate simulation at each state point and converges slowly for such species.

In bulk, the solute distribution is Poisson-like and strongly skewed, while that of the abundantly present solvent is nearly Gaussian.
The grand-potential minimum, which marks the most probable particle number, therefore coincides with the ensemble average for the solvent, but lies systematically below the equilibrium occupancy $\langle N_\mathrm{B}\rangle$ of the solute and is zero whenever $\langle N_\mathrm{B}\rangle < 1$, which would wrongly suggest that no solute is dissolved.
This distinction is expected whenever a dilute species has a mean number of order one or less in the volume considered, either because of poor solubility, low reservoir pressure, or a small volume such as a nanopore.

Under slit-pore confinement, the wall--solvent attraction controls the oversolubility at every pore width studied, whereas the wall--solute attraction changes it by about $10\,\%$ at most.
The solute excess builds up where the wall excludes the solvent, regardless of whether the wall attracts the solute, as the nearly coincident density profiles of cases I and II show.
Therefore, the wall--solvent interaction sets most of the enhancement, which isolates and quantifies the role of the liquid--solid interaction proposed by Coasne et al.~\cite{Ho2013, Coasne2019}.

This dominant wall--solvent contribution is consistent with simulations of \ce{CO2} in water-filled silica and clay nanopores, which find the largest oversolubility for hydrophobic walls~\cite{Xie2026, Li2020} and under-solubility for strongly hydrophilic ones~\cite{Li2020}.
Our results suggest that these trends arise mainly from the affinity of the wall for water (solvent) rather than from the adsorption of \ce{CO2} (solute).

In a forthcoming study, we apply the method to an all-atom model of \ce{CO2} in water confined in carbon slit pores, to test whether the wall--water affinity also governs the oversolubility of \ce{CO2} in real nanopores.
The method can also be applied to confined electrolyte solutions, where dissolved ions can increase or decrease the solubility of \ce{CO2} depending on where they adsorb~\cite{Ali2021}; the same implementation has already been used for electrolytes in inert and conducting nanopores~\cite{Stark2026}.
Together, these developments would provide a route for screening nanoporous materials for carbon capture, catalysis, and gas storage~\cite{Mathur2026, Micari2025, Zhang2024, Zhang2024_MOF}.

%% file: sections/appendix.tex
\appendix

\section{Error estimation}
\label{app:errors}

\subsection{Intrinsic error in the density of states}
\label{sec:error_estimation}

Our implementation estimates $\ln\Omega$ directly, so we expect the $\sqrt{\ln f}$ scaling of Zhou and Bhatt~\cite{Zhou2005} to hold, with a non-universal prefactor set by sampling correlations and implementation details.

Since the exact density of states is unknown, its intrinsic error cannot be evaluated directly.
Instead, we monitor the change of the estimated $\ln\Omega$ between two successive Wang--Landau iterations $k-1$ and $k$.
For a macrostate $i\in\{1,\ldots,M\}$, with $M$ the number of macrostates, we define
\begin{align}
    d_i^{(k)} &= \ln\Omega_i^{(k)} - \ln\Omega_i^{(k-1)},
    \qquad
    c_k = \frac{1}{M}\sum_{i=1}^{M}d_i^{(k)},
    \nonumber\\
    D_k &= \left[\frac{1}{M}\sum_{i=1}^{M}\left(d_i^{(k)}-c_k\right)^2\right]^{1/2},
\end{align}
so that $D_k$ is the root-mean-square change in the shape of the $\ln\Omega$ profile, with the arbitrary additive shift $c_k$ removed.
If the finite modification factor dominates the final iterations, $D_k$ is expected to follow
\begin{equation}
    D_k = C_{\ln\Omega}\sqrt{\ln f(k,\alpha)},
    \qquad
    f(k,\alpha) = \exp\left(\frac{\alpha}{2^k}\right),
\end{equation}
where $C_{\ln\Omega}$ is a system- and implementation-dependent prefactor and $\alpha=1$.
\begin{figure}
    \centering
    \includegraphics[width=0.8\linewidth]{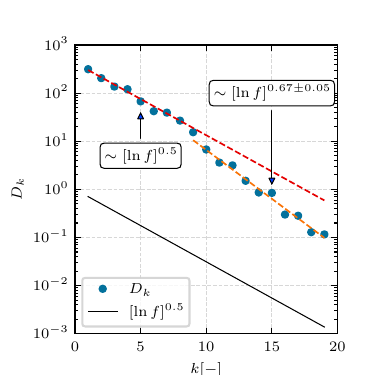}
    \caption{Root-mean-square variation $D_k$ of the shape of the log-density-of-states profile of the bulk system as a function of the iteration counter $k$. The solid black line shows $\sqrt{\ln f(k)}$.}
    \label{fig:RMS_variation_lnOmega}
\end{figure}

For the bulk system (Fig.~\ref{fig:RMS_variation_lnOmega}), $D_k$ follows $\sqrt{\ln f}$ up to $k=8$ and then decays faster, as $(\ln f)^{n}$ with $n=0.67\pm0.05$.
This faster decay does not indicate faster convergence.
Since $D_k$ compares successive estimates, correlations between consecutive iterations suppress it; the large minimum number of visits per stage ($\varepsilon=1000$) and a saturation of the late-iteration updates can contribute as well.
All sampling runs start from the same initial configuration.
$D_k$ is therefore used as a convergence diagnostic for $\ln\Omega$, not as a direct measurement of its absolute uncertainty.
Under confinement, the late-iteration exponent is unchanged, whereas the crossover moves to earlier iterations as the pore narrows, down to $k=2$ for the narrowest pore studied.

\begin{figure*}
    \centering
    \includegraphics[width=0.8\linewidth]{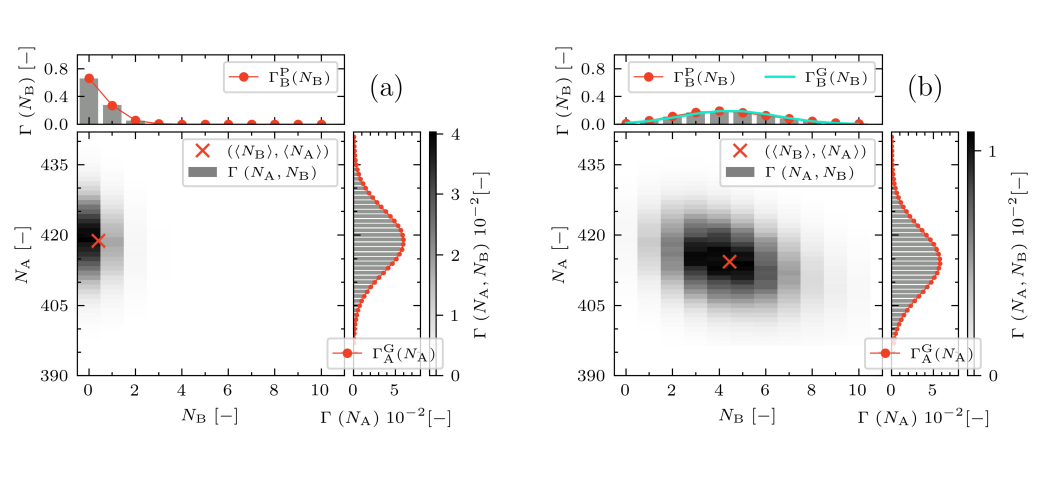}
    \caption{Joint particle-number distribution of solute (horizontal axis) and solvent (vertical axis) for case II in a pore of width $h=8\,\sigma$, at $\mu_\mathrm{A} = -1.1\,\epsilon$ and (a) $\mu_\mathrm{B} = -3.8\,\epsilon$, (b) $\mu_\mathrm{B} = -1.5\,\epsilon$. The top and right panels show the marginal distributions of the solute and the solvent. Red curves: Poisson distribution $\Gamma^{\mathrm{P}}_{\mathrm{B}}(N_\mathrm{B})$ for the solute and Gaussian distribution $\Gamma^{\mathrm{G}}_{\mathrm{A}}(N_\mathrm{A})$ for the solvent. Green curve in (b): Gaussian distribution $\Gamma^{\mathrm{G}}_{\mathrm{B}}(N_\mathrm{B})$, which the solute distribution approaches at larger $\langle N_\mathrm{B}\rangle$.}
   \label{fig:confined_system}
\end{figure*}

\begin{figure*}
    \centering
    \subfloat[A slit pore of width $h = 5\,\sigma$.\label{fig:mf_5}]{
        \includegraphics[width=0.4\linewidth]{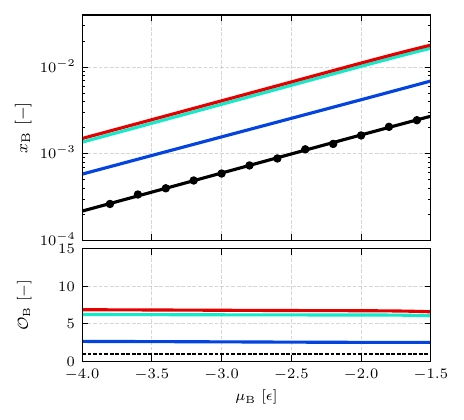}
    }\hfill
    \subfloat[A slit pore of width $h = 8\,\sigma$.\label{fig:mf_8}]{
        \includegraphics[width=0.4\linewidth]{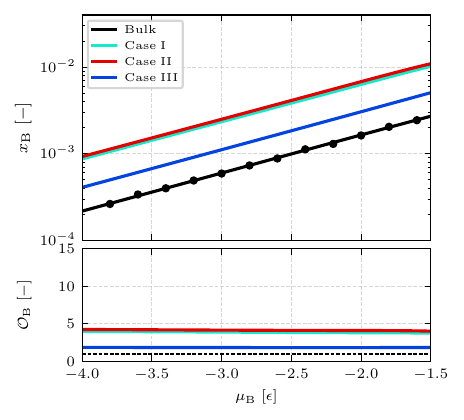}
    }
   \caption{Solute mole fraction $x_\mathrm{B}$ (top panels) and solute oversolubility $\mathcal{O}_\mathrm{B}$ (bottom panels) as functions of the solute chemical potential $\mu_\mathrm{B}$, for pore widths $h = 5$ and $8\,\sigma$ and the three cases of Table~\ref{tab:cases}. The black dashed line marks $\mathcal{O}_{\mathrm{B}}=1$, where the confined solute mole fraction equals its bulk value.}
    \label{fig:solute_mole_fraction}
\end{figure*}

\subsection{Window-stitching error}
\label{sec:error_stitching}

A second source of error arises from combining the sampling windows.
Each window $m\in\{1,\ldots,W\}$ yields $\ln\Omega_m(\mathbf{N})$ up to an unknown additive shift $\Delta_m$, where $\mathbf{N}=(N_{\mathrm{A}},N_{\mathrm{B}})$ denotes a macrostate.
For each sampled macrostate $\mathbf{N}_i$, we set $C_{im}=1$ if $\mathbf{N}_i$ belongs to window $m$ and $C_{im}=0$ otherwise, and $n_i=\sum_{m=1}^{W}C_{im}$ is the number of windows covering $\mathbf{N}_i$.
The stitched density of states is the mean of the shifted window values,
\begin{equation}
\label{eq:dos_mean}
\ln\Omega(\mathbf{N}_i)
= \frac{1}{n_i}
\sum_{m=1}^{W} C_{im}
\left[\ln\Omega_m(\mathbf{N}_i) + \Delta_m\right],
\end{equation}
and the shifts $\boldsymbol{\Delta}=(\Delta_1,\ldots,\Delta_W)^T$ minimize the squared deviations of the shifted window values from this mean,
\begin{equation}
J(\boldsymbol{\Delta}) =
\sum_i \sum_{m=1}^{W} C_{im}
\left[
\ln\Omega_m(\mathbf{N}_i) + \Delta_m - \ln\Omega(\mathbf{N}_i)
\right]^2,
\label{eq:dos_objective}
\end{equation}
a linear least-squares problem that we solve with SciPy\cite{Virtanen2020}.
The stitching error is the sample standard deviation of the shifted window values at each macrostate,
\begin{align}
\sigma^2_{\mathrm{stitching}}(\mathbf{N}_i) &=
\frac{1}{n_i-1}
\sum_{m=1}^{W} C_{im}
\nonumber\\
&\quad\times\left[
  \ln\Omega_m(\mathbf{N}_i) + \Delta_m - \ln\Omega(\mathbf{N}_i)
\right]^2.
\label{eq:agreement_error}
\end{align}
It vanishes when all overlapping windows agree and is undefined for macrostates covered by a single window.

\subsection{Error of statistical properties}
\label{sec:error_estimation_of_stat_props}

We assign each macrostate $(N_{\mathrm{A},k}, N_{\mathrm{B},l})$ an uncertainty $\sigma_{\ln\Omega_{kl}}$ of $\ln\Omega_{kl}$, taken from $\sigma_{\ln\Omega}$ defined in Sec.~\ref{sec:methods}, treat these uncertainties as independent, and regard the chemical potentials as exact.
Independence has to be assumed for a single sampling run; repeated runs would give access to the covariances.
With $\omega_{kl} = \beta\left(N_{\mathrm{A},k}\mu_{\mathrm{A}} + N_{\mathrm{B},l}\mu_{\mathrm{B}}\right)$, the joint distribution is $\Gamma_{kl} = e^{\ln\Omega_{kl}+\omega_{kl}}/\Xi$, and the ensemble average of either species, $\langle N\rangle = \sum_{kl} N_{kl}\Gamma_{kl}$, where $N_{kl}$ is the particle number of that species in macrostate $(k,l)$, changes as
\begin{equation}
    \frac{\partial\langle N\rangle}{\partial\ln\Omega_{kl}} = \Gamma_{kl}\left(N_{kl}-\langle N\rangle\right).
\end{equation}
First-order error propagation\cite{Bevington2003} then gives
\begin{equation}
    \sigma_{\langle N\rangle}
    =
    \left[
        \sum_{k,l}
        \sigma^2_{\ln\Omega_{kl}}\,
        \Gamma_{kl}^2
        \left(N_{kl}-\langle N\rangle\right)^2
    \right]^{1/2}.
\end{equation}
The errors of $x_{\mathrm{B}}$, $\mathcal{O}_{\mathrm{B}}$, and the $\mu_{\mathrm{B}}$-averaged $\langle\mathcal{O}_{\mathrm{B}}\rangle$ follow by the same first-order propagation.
We neglect the covariance of $\langle N_{\mathrm{A}}\rangle$ and $\langle N_{\mathrm{B}}\rangle$, since the relative error of the abundant solvent is much smaller than that of the solute.
Because $D_k$ measures the change between successive iterations and the macrostate errors are treated as independent, the resulting error bars are order-of-magnitude estimates.

\section{Particle-number distributions and oversolubility in wider pores}
\label{app:wider}

Figure~\ref{fig:confined_system} shows the joint particle-number distribution in a slit pore, and Fig.~\ref{fig:solute_mole_fraction} shows the solute mole fraction and oversolubility in the two wider pores.